\documentclass[12pt]{article}
\usepackage[cp866]{inputenc}
\usepackage[english]{babel}
\usepackage{graphicx}
\usepackage{subfig}
\usepackage{extsizes}
\usepackage{epsfig}
\usepackage[numbers,sort&compress]{natbib}
\usepackage{citehack}
\usepackage{float}
\usepackage{ams math}
\usepackage{amssymb}
\usepackage{multicol}
\usepackage{amscd}
\usepackage{verbatim}
\usepackage{booktabs}
\usepackage{enumitem}
\usepackage{tikz}
\newcommand{\eps}{\varepsilon}

\makeatletter \@addtoreset{equation}{section} \makeatother
\allowdisplaybreaks[1]

\begin{document}
      \title{Unusual Properties of Adiabatic Invariance in a Billiard Model Related to the Adiabatic Piston Problem}
      \date{}
      \author{Joshua Skinner, Anatoly Neishtadt} 
      \maketitle
    \begin{abstract}
    We consider the motion of two massive particles along a straight line. A lighter particle bounces back and forth between a heavier particle and a stationary wall, with all collisions being ideally elastic. It is known that if the lighter particle moves much faster than the heavier one, and the kinetic energies of the particles are of the same order, then the product of the speed of the lighter particle and the distance between the heavier particle and the wall is an adiabatic invariant: its value remains approximately constant over a long period.
We show that the value of this adiabatic invariant, calculated at the collisions of the lighter particle with the wall, is a constant of motion (i.e., {an exact adiabatic invariant}). On the other hand, the value of this adiabatic invariant at the collisions between the particles slowly linearly in time  decays with each collision.

The model we consider is a highly simplified version of the classical adiabatic piston problem, where the lighter particle represents a gas particle, and the heavier particle represents the piston.

       \end{abstract}

\section {Statement of the problem } 
\begin{figure}[H]
    \centering
\begin{tikzpicture}
 \draw (8,0) rectangle (8.5,4);
    \draw (2,0) -- (2,4);
    \draw (5,2) circle (0.1cm);
    \draw (2,4) -- (1,3);
    \draw (2,3) -- (1,2);
    \draw (2,2) -- (1,1);
    \draw (2,1) -- (1,0);
    \draw (3.5,2) -- (4.2,2);
    \draw (4,1.8) -- (4.2,2) -- (4,2.2);
    \draw (8.8,2) -- (9.8,2);
    \draw (4,1.8) -- (4.2,2) -- (4,2.2);
    \draw (9,2.2) -- (8.8,2) -- (9,1.8);
\end{tikzpicture}
\caption{Diagram displaying the interaction between the piston and gas particle}
\label{AdiaPistonDiagram}
\end{figure}
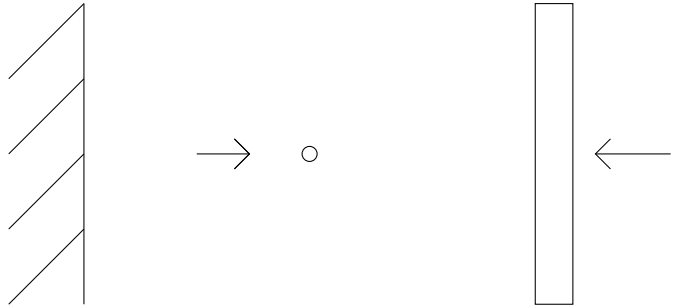

Consider the motion of two particles along a straight line. Denote $m$ and $M$ masses of these particles. We assume $m<M$.  We examine the case where the lighter particle bounces back and forth between a stationary wall and the heavier particle, undergoing elastic collisions with both. This model was considered in \cite{galperin, gorelyshev}.
 We refer to the lighter particle as the ``gas particle" and the heavier particle as the ``piston", due to the connection of this model to the classical adiabatic piston problem \cite{gruber} (Fig.\ref{AdiaPistonDiagram}).

Take stationary wall as the origin of the coordinate line. Let the particles move in  the positive half-line. Denote $x$ and $X$ coordinates of the gas particle and the piston, respectively.   Thus, $0\le x\le X$, and $X$ is the distance between the stationary wall and  the piston.

Denote $v$ and $V$ velocities of the gas particle and the piston, respectively. We assume that $|v|> |V|$. Thus, the gas particle moves  faster than the piston. 
Denote $I=|v|X$. It is known that if $m\ll M$ and $|v|\gg |V|$ then $I$ is an adiabatic invariant,  meaning its value remains approximately constant over long time intervals. We aim to study the behavior of $I$ in more details.

 \section{Values of adiabatic invariant at collisions with the stationary wall}
 Let the gas particle collide with the stationary wall at a certain moment of time. 
 Denote by $v_0$ the velocity of the gas particle immediately after this collision, $v_0>0$.  Denote by $V_0, X_0$ the piston's velocity and coordinate (i.e. the distance between the stationary wall and the piston) at the time of this collision. The value of the adiabatic invariant $I$ at this collision is $I_0=v_0X_0$.
 
 The time interval between this collision and  the subsequent  collision of the gas particle with the piston is
 $$
\Delta t_1=\frac{X_0}{v_0-V_0}.
 $$ 
 The distance between the stationary wall and the piston at the time of the collision of the gas particle with the piston  is
 $$
 X_1=X_0+V_0\Delta t_1= \frac{X_0}{v_0-V_0}v_0.
 $$
 The velocities of the gas particle and the piston before this collision are still $v_0, V_0$. Denote by $v_1, V_1$ the velocities of the gas particle and the piston immediately after this collision. According to the standard formulas  for ideally elastic collisions:
 \begin{equation}
 \begin{aligned}
 \label{ideal}
 &v_1=\frac{(m-M)v_0+2MV_0}{M+m},     \\
 &V_1=\frac{(M-m)V_0+2mv_0}{M+m}.
 \end{aligned}
 \end{equation}
 The gas particle will then collide with the stationary wall after the time interval 
 $$
  \Delta t_2=\frac{X_1}{-v_1}.
 $$ 
 The distance between the stationary wall and the piston at this collision is
 $$
 X_2=X_1+V_1 \Delta t_2=X_1+\frac{V_1X_1}{-v_1}=X_1\frac{v_1-V_1}{v_1}.
 $$
 The value of the adiabatic invariant $I$ at this collision is $I_2=-v_1X_2$. We have
  \begin{equation*}
 \begin{aligned}
I_2= -v_1X_2=-v_1X_1\frac{v_1-V_1}{v_1}=-X_1(v_1-V_1)=-\frac{X_0}{v_0-V_0}v_0(v_1-V_1)=-I_0\frac{v_1-V_1}{v_0-V_0}.
  \end{aligned}
 \end{equation*}
 According to (\ref{ideal})
 $$
 v_1-V_1=\frac{(m-M)v_0+2MV_0}{M+m}-     
\frac{(M-m)V_0+2mv_0}{M+m}=-(v_0-V_0).
 $$ 
 Thus, $I_2=I_0$. One can repeat these calculations for any two consecutive collisions with the stationary wall. Therefore, the adiabatic invariant $I$  retains the same value at any collision with the stationary wall. In the terminology of \cite{veselov}, the value of $I$ at collisions with the stationary wall is {an exact adiabatic invariant}.
    
 \section{Values of adiabatic invariant at collisions with the piston}
 \label{Values of adiabatic invariant at collisions with the piston}
 Let the gas particle collide with the piston  in some moment of time. 
 Denote by $v_1, V_1$  velocities of the gas particle and the piston immediately after this collision, $v_1<0$.  Denote by $X_1$ the piston's coordinate (i.e., the distance between the stationary wall and the piston) at the time of this collision. The value of the adiabatic invariant $I$ at this collision is $I_1=-v_1X_1$.
 
 The gas particle will collide with the piston again after a time $ \Delta t_3$ which can be found from  the equation
 $$
 X_1+V_1 \Delta t_3=(-v_1) \Delta t_3 -X_1.
 $$ 
 Thus,
 $$
  \Delta t_3=-\frac{2X_1}{V_1+v_1}.
 $$
 The distance between the stationary wall and the piston at this collision is
 $$
 X_3=X_1+V_1 \Delta t_3= X_1-V_1 \frac{2X_1}{V_1+v_1}.
 $$
 The velocity of the gas particle just after this collision is
 $$
 v_3=\frac{(m-M)(-v_1)+2MV_1}{M+m}. 
 $$
 We assume that $v_3<0$. Thus, the particle moves to the stationary wall after this collission.  The value of the adiabatic invariant $I$ just after this collision is $I_3=-v_3X_3$. We have
  \begin{equation*}
 \begin{aligned}
I_3=-v_3X_3= - \frac{(m-M)(-v_1)+2MV_1}{M+m} \left(X_1-V_1 \frac{2X_1}{V_1+v_1} \right).    \end{aligned}
 \end{equation*}
 Using Maple, this can be simplified to
 $$
 I_3=I_1+\frac{2X_1(MV_1^2+mv_1^2)}{(M+m)(V_1+v_1)}.
 $$
 Thus, change of the value of the adiabatic invariant $I$ calculated at the times of collisions of the gas particle and the piston is 
 $$
 \Delta I_3=\frac{2X_1(MV_1^2+mv_1^2)}{(M+m)(V_1+v_1)}.
 $$
 Calculate the rate of this change
 $$
 k=\frac{\Delta I_3}{\Delta t_3}=-\frac{(MV_1^2+mv_1^2)}{(M+m)} 
 $$
 But $(MV_1^2+mv_1^2)/2$ is the kinetic energy of the system. It does not change at collisions. Thus, the value $k$ is the same for all collisions. Therefore, the value of this adiabatic invariant at the collisions between the particles linearly in time  decays with each collision.  However, in an adiabatic situation, where $|v_1|\gg|V_1|$ and $m\ll M$, this decay occurs very slowly. Indeed, the case of interest is when the kinetic energies of the gas particle and the piston are of the same order and are of order 1: $MV_1^2\sim mv_1^2\sim 1$. We can take 
$v_1\sim 1, m\sim1, V_1\sim \eps \ll 1, M\sim 1/\eps^2, X\sim 1$.  Then $I_1\sim 1$ and $k \sim \eps^2$.

 \section{Numerical Illustrations}
 
 This section is dedicated to numerical illustrations of the dynamics in the considered problem. We assume the masses of the particle and piston to be $m=1$ and $M=10000$, respectively.
The initial velocities of the particle and piston are $v_0=1$ and $V_0=-0.01$. Thus, the piston initially moves towards the stationary wall. The initial positions of the particle and piston are $x_0=0$ and $X_0=1$.
\medskip

\begin{figure}[H]
     \centering
     \includegraphics[width=0.7\linewidth]{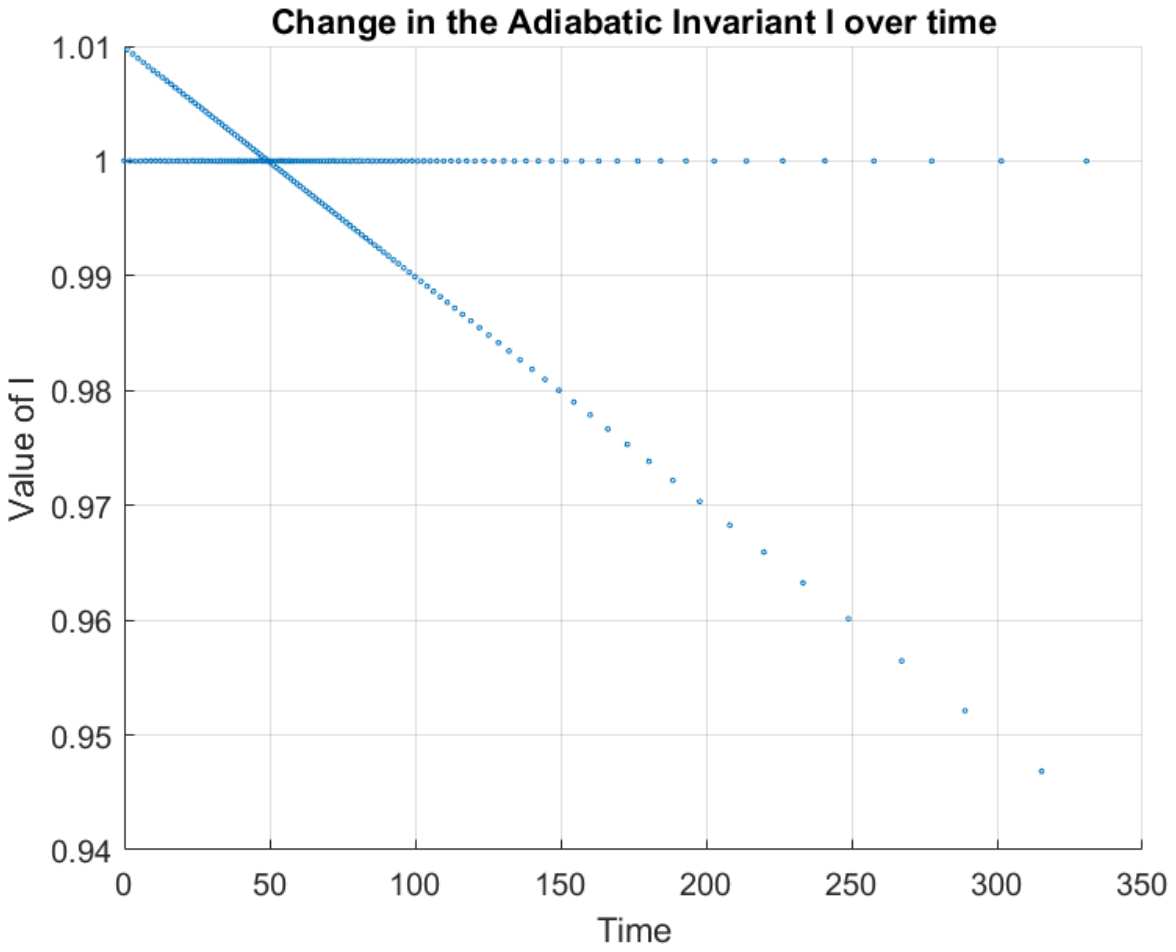}
     \caption{Graph depicting the change in the adiabatic invariant \textit{I} as the system evolves in time}
     \label{fig: I_t}
 \end{figure}

\begin{figure}[H]
     \centering
     \includegraphics[width=0.7\linewidth]{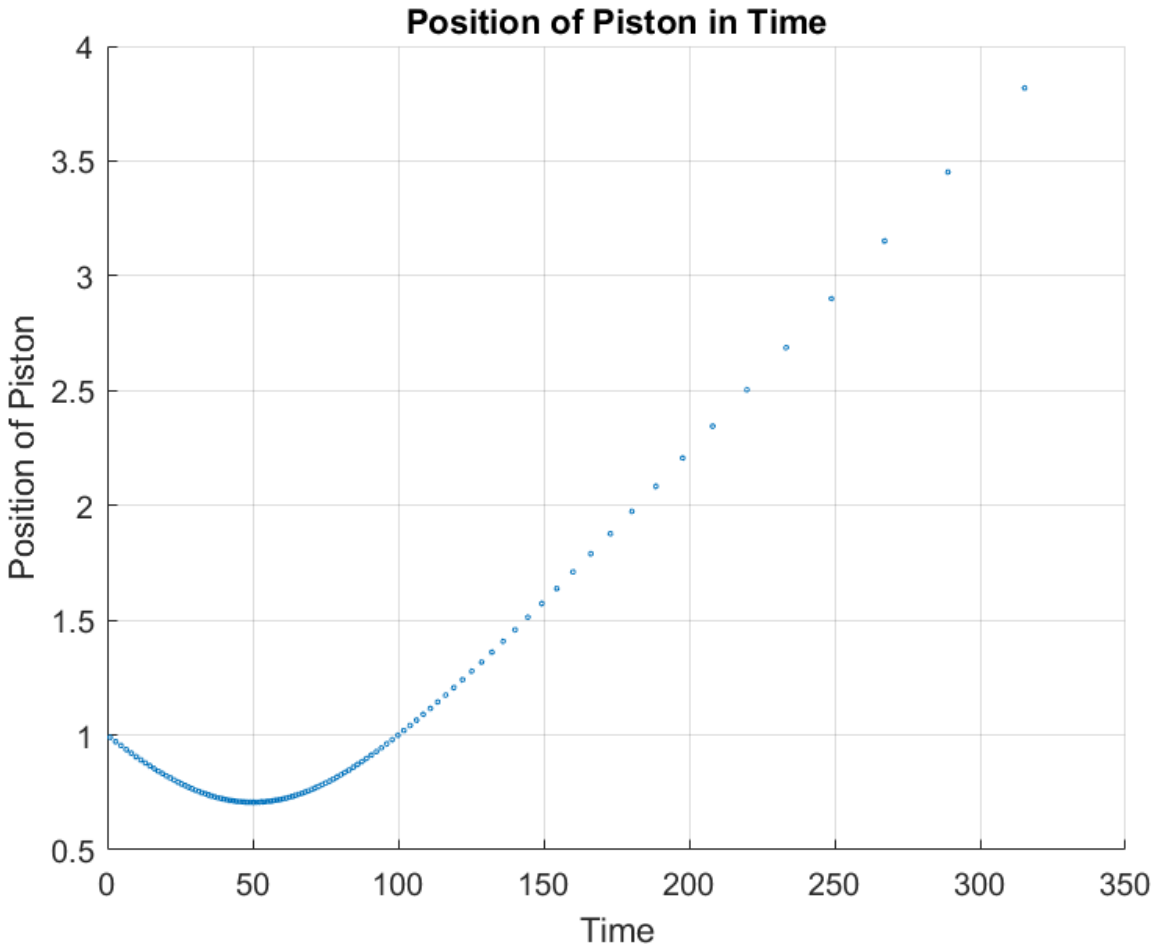}
     \caption{Graph depicting the position of the piston as the system evolves in time}
     \label{fig: X_t}
 \end{figure}

\begin{figure}[H]
     \centering
     \includegraphics[width=0.7\linewidth]{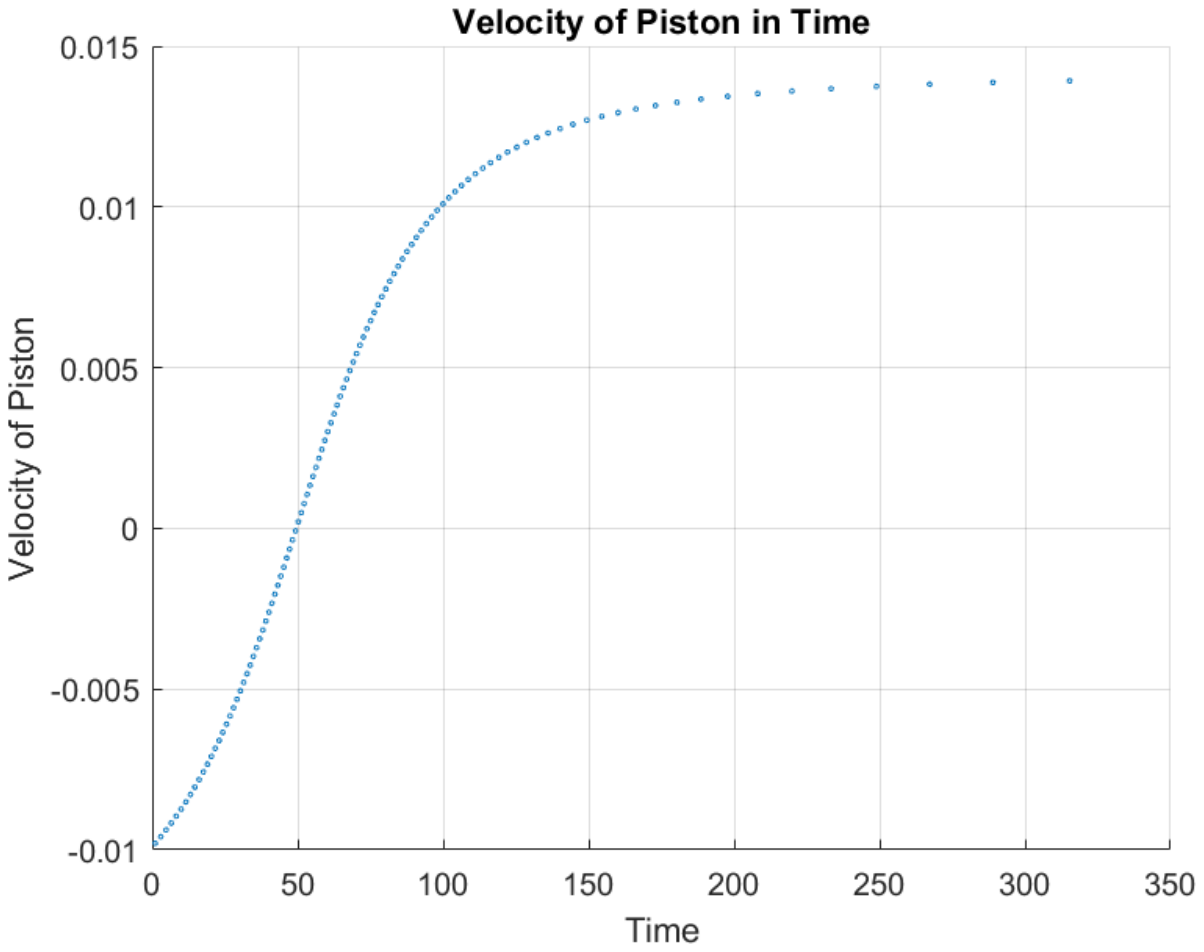}
     \caption{Graph depicting the velocity of the piston as the system evolves in time}
     \label{fig: V_t}
 \end{figure}

Fig. \ref{fig: I_t} illustrates the main finding of this paper: the values of the adiabatic invariant $I$
at the collisions between the particle and the stationary wall remain constant (represented by the horizontal line in Fig. \ref{fig: I_t}), while the values at the collisions with the piston decrease linearly in time (represented by the inclined line in Fig. \ref{fig: I_t}).
 As seen in Figs. \ref{fig: X_t} and \ref{fig: V_t}, the piston initially moves towards the stationary wall. Due to the approximate conservation of the adiabatic invariant, as the distance between the piston and the stationary wall decreases, the particle's velocity increases, resulting in a rise in pressure on the piston. Consequently, the piston eventually comes to a stop and begins moving in the opposite direction, with its velocity increasing after each subsequent collision.
   
\begin{figure}[H]
     \centering
     \includegraphics[width=0.7\linewidth]{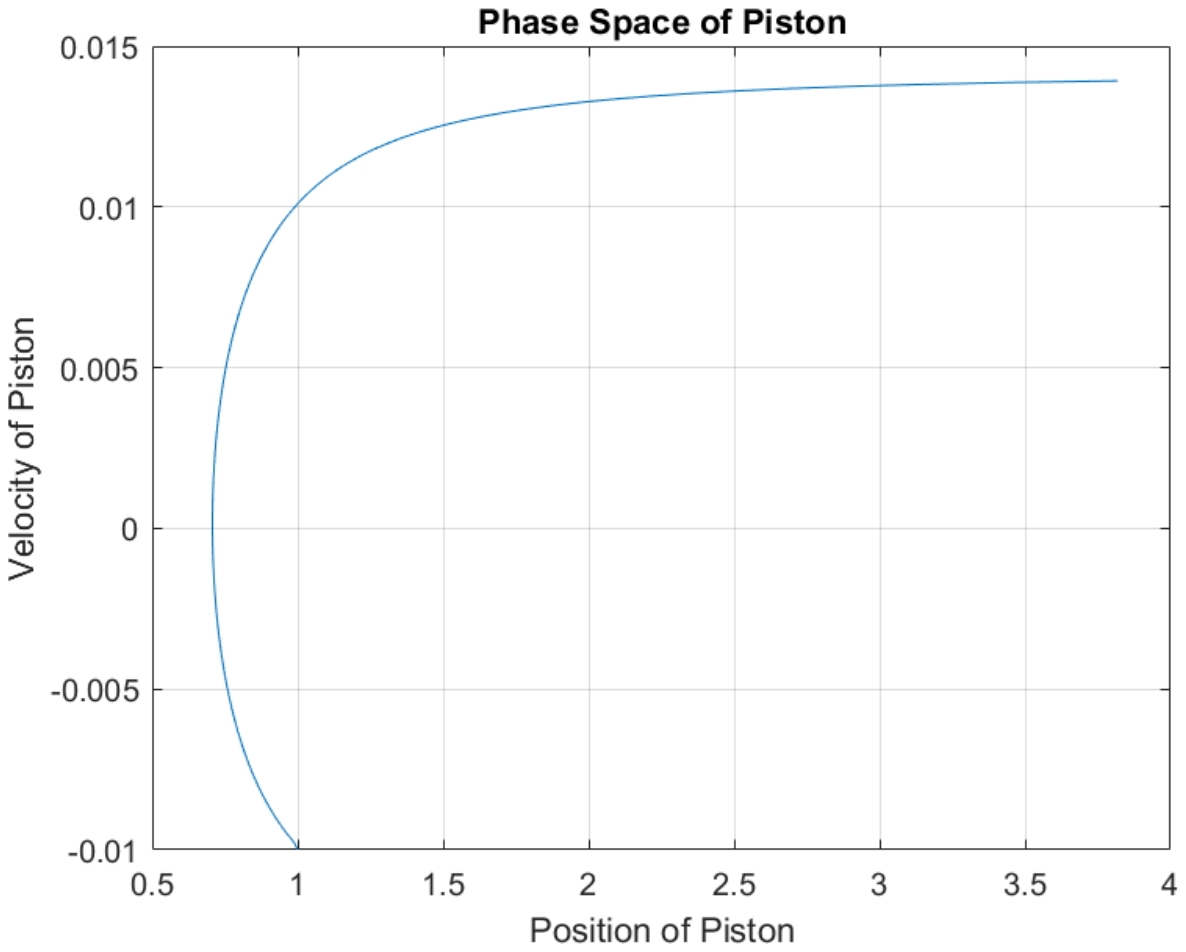}
     \caption{Trajectory of the piston in the phase plane}
     \label{fig: V_X}
 \end{figure}

 Figure \ref{fig: V_X} describes the evolution of the pistons velocity and position after each collision with the piston. As can also be seen in figures \ref{fig: X_t} and \ref{fig: V_t}, the piston travels towards the stationary wall until a turning point is reached, at which the piston halts and travels away from the stationary wall, increasing in speed after further collisions. The piston velocity begins to plateau as the piston moves further away due to the particle having less kinetic energy to transfer to the piston as they collide.

 \section{Conclusions}
 The  problem considered is a slow-fast Hamiltonian system with two degrees of freedom and  elastic collisions. In this system, the fast degree of freedom corresponds to the gas particle, while the slow degree of freedom corresponds to the piston. In such systems, the expected behavior of the adiabatic invariant is that its value oscillates with a small amplitude around some constant value. However, in the problem examined, the behavior of the adiabatic invariant is somewhat unusual. Specifically, its value at the collision of the gas particle with the stationary wall remains constant, while its value at the collision of the gas particle with the piston linearly in time decays from collision to collision.

\medskip
 
 The authors would like to thank A.P. Veselov for useful discussions.

\vskip 5mm

\noindent Joshua Skinner

\noindent {\small Department of Mathematical Sciences}

\noindent {\small Loughborough University, Loughborough LE11 3TU, United Kingdom}

\noindent {\footnotesize{E-mail: skinnerjoshua515@gmail.com}}

\vskip 3mm

\noindent Anatoly Neishtadt

\noindent {\small Department of Mathematical Sciences}

\noindent {\small Loughborough University, Loughborough LE11 3TU, United Kingdom}

\noindent {\footnotesize{E-mail: a.neishtadt@lboro.ac.uk}}


\newpage

\begin{thebibliography}{99}

\bibitem{galperin} Galperin G.A. Playing pool with $\pi$ (the number $\pi$ from a billiard point of view). Regul. Chaotic Dyn.,  {\bf 8},  375--394 (2003)

\bibitem{gorelyshev} Gorelyshev I.V. On the full number of collisions in certain one-dimensional billiard problems.
Regul. Chaotic Dyn.,  {\bf 11},  61--66 (2006)

\bibitem{gruber} Gruber C., Lesne A. Adiabatic Piston. In: Encyclopedia of Mathematical Physics, 160--173, Elsevier (2006)
 

\bibitem{veselov} Veselov A.P. A few things I learnt from J\"urgen Moser.
Regul. Chaotic Dyn.,  {\bf 13},  515--524  (2008)






\end{thebibliography}
\end{document}